\documentclass[prb,showpacs,preprintnumbers,amsfonts,amssymb,amsmath,floats,twocolumn,superscriptaddress,aps]{revtex4}

\usepackage{graphicx}
\usepackage{dcolumn}
\usepackage{bm}
\usepackage[final,dvips]{epsfig}
\usepackage{bm}
\usepackage{color}
\usepackage{soul}
\usepackage{subfigure}

\newcommand{\ff}[1]{{\boldsymbol #1}}
\newcommand{\ca}[1]{{\cal #1}}

\begin{document}

\title{Crossover from conventional to inverse indirect magnetic exchange in the depleted Anderson lattice}

\author{Maximilian W. Aulbach}
\affiliation{I. Institute for Theoretical Physics, University of Hamburg, Jungiusstra\ss{}e 9, 20355 Hamburg, Germany}
\author{Irakli Titvinidze}
\affiliation{I. Institute for Theoretical Physics, University of Hamburg, Jungiusstra\ss{}e 9, 20355 Hamburg, Germany}
\affiliation{Institute of Theoretical and Computational Physics, Graz University of Technology, Petersgasse 16, 8010 Graz, Austria}
\author{Michael Potthoff}
\affiliation{I. Institute for Theoretical Physics, University of Hamburg, Jungiusstra\ss{}e 9, 20355 Hamburg, Germany}

\begin{abstract}
We investigate the finite-temperature properties of an Anderson lattice with regularly depleted impurities.
The physics of this model is ruled by two different magnetic exchange mechanisms: 
conventional Ruderman-Kittel-Kasuya-Yosida (RKKY) interaction at weak hybridization strength $V$ and a novel inverse indirect magnetic exchange (IIME) at strong $V$, both favoring a ferromagnetic ground state. 
The stability of ferromagnetic order against thermal fluctuations is systematically studied by static mean-field theory for an effective low-energy spin-only model emerging perturbatively in the strong-coupling limit as well as by dynamical mean-field theory for the full model. 
The Curie temperature is found at a maximum for a half-filled conduction band and at intermediate hybridization strengths in the crossover regime between RKKY and IIME.
\end{abstract}
 


\maketitle

\section{Introduction}

As has been pointed out by Nozi\`eres, \cite{Noz74,Noz76,NB80} the presence of a correlated impurity in an {\em a priori} uncorrelated metal introduces {\em effective} interactions among the conduction electrons. 
The range of these interactions decisively depends on the strength of the impurity-host coupling. 
Consider the case of an Anderson impurity, \cite{And61} 
\begin{eqnarray}
H
&=&
- t \sum_{\langle i,j \rangle} \sum_{\sigma=\uparrow, \downarrow} c^\dagger_{i\sigma} c_{j\sigma}
+ V \sum_{\sigma=\uparrow, \downarrow} \left(c^\dagger_{i_{0} \sigma} f_{\sigma} + \mbox{h.c.} \right)  
\nonumber \\
&+&
U (f_{\uparrow}^{\dagger} f_{\uparrow} - 1/2) (f_{\downarrow}^{\dagger} f_{\downarrow} - 1/2)
\, ,
\label{eq:and}
\end{eqnarray}
with annihilators $c_{i\sigma}, f_{\sigma}$ referring to local conduction-electron and impurity orbitals, respectively.
For the case of a Hubbard interaction $U$ and a local hybridization $V$ much stronger than the nearest-neighbor conduction-electron hopping $t$, an effective Hamiltonian with an almost local interaction characterizing the low-energy physics of the conduction-electron system can be derived explicitly. \cite{TSP15}
This is achieved by means of degenerate fourth-order perturbation theory in the hopping terms which connect the neighbouring conduction-electron sites to the site $i_{0}$ where the impurity is coupled to. 
To leading order, the effective model is given by 
\begin{equation}
H_{\rm eff}
=
- t \sum_{\langle i,j \rangle}^{i,j \ne i_{0}} \sum_{\sigma=\uparrow, \downarrow} c^\dagger_{i\sigma} c_{j\sigma}
-
\frac{z^{2} \alpha}{3}
\ff S_{\rm bond}^{2}
\, ,
\label{eq:effimp}
\end{equation}
where $z$ is the coordination number of the lattice, where
\begin{equation}
  \alpha = t^{4} \frac{U^{3} + 48 U V^{2}}{24 V^{6}}
\label{eq:alpha}
\end{equation}
is the effective interaction strength, and where $\ff S_{\rm bond}$ is the spin-operator referring to the ``bonding'' symmetric linear superposition of the $z$ orbitals neighbouring $i_{0}$ (see Ref.\ \onlinecite{TSP15} for details).

There are three different energy scales to be considered:
(i) Local singlet formation at $i_{0}$ takes place on the high-energy scale $\sim U, V$. 
While this singlet may be called a local {\em Kondo} singlet, its binding energy scales linearly with $V$ for strong $V$. 
This is opposed to the weak-coupling limit $V \to 0$ (with $U\gg t$ fixed) where it is exponentially small and where the low-energy physics is dictated by a single Kondo scale. \cite{Wil75,Hew93}
(ii) On an energy scale $\sim t$, conduction electrons scatter at the local Kondo singlet. 
This scattering effect is already included at zeroth order in the perturbative expansion and is formally described by excluding the site $i_{0}$ from the summation in the first term of the effective Hamiltonian in Eq.\ (\ref{eq:effimp}).
(iii) The first non-trivial effect takes place at fourth order. 
An effective interaction among the conduction electrons in the immediate vicinity of the impurity emerges which is mediated by virtual excitations of the local Kondo singlet. 
This happens on the lowest energy scale given by the effective coupling constant $\alpha$ in the second term of Eq.\ (\ref{eq:effimp}).

\begin{figure}[t]
  \centerline{\includegraphics[width=0.4\textwidth]{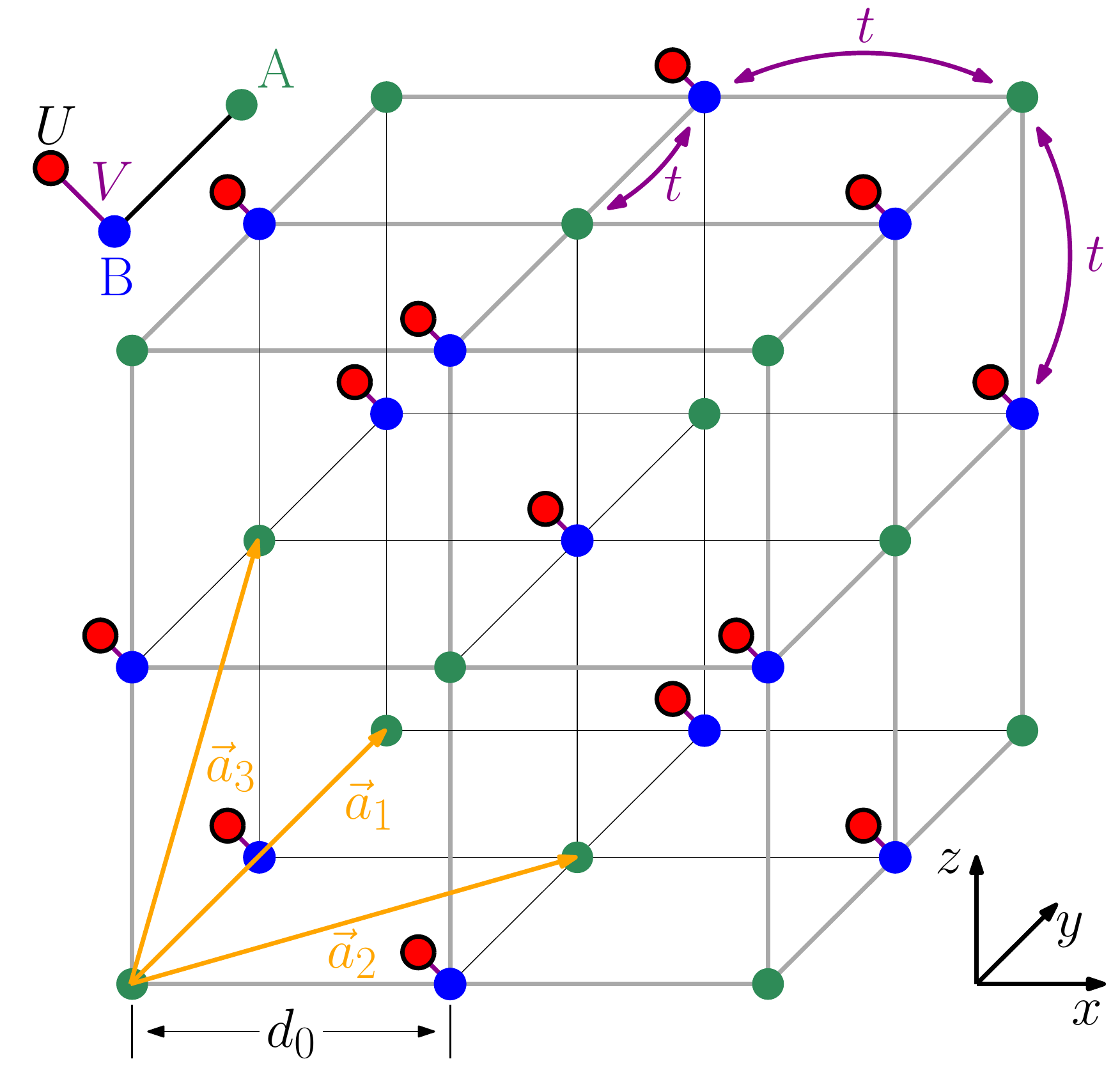}}
  \caption{(Color online) 
Depleted periodic Anderson model with $R=L/2$ impurities on a $D=3$-dimensional simple-cubic lattice with $L$ sites ($L \to \infty$).
Correlated impurity sites (red) with on-site Hubbard interaction $U$ are coupled via a hybridization of strength $V$ to the B sites (blue) of the bipartite lattice. 
For strong $U, V \gg t$, local Kondo singlets are formed on the half-filled ``dimers'' consisting of impurity and B sites (if the total electron number $N$ satisfies $2R\le N \le 4R$) and strongly confine the motion of the excess conduction electrons on the A-sublattice sites (green). 
Virtual excitations of the local Kondo singlets induce an effective interaction of the conduction electrons on A sites.
}
\label{fig:3d}
\end{figure}

A fundamentally interesting question is whether the emergent {\em effective} interaction among the {\em a priori} uncorrelated conduction electrons can give rise to collective phenomena. 
This may be expected for a lattice variant of the model, i.e., for a system with a thermodynamically relevant concentration of impurities. 
The extreme case is a periodic Anderson model with a depleted system of ``impurities'' placed at every second site, i.e., on the B sites of a bipartite lattice consisting of sublattices A and B.
Fig.\ \ref{fig:3d} displays an example for the $D=3$ dimensional simple-cubic lattice.
We consider a model with $L$ sites ($L\to \infty$ in the thermodynamical limit) and $R=L/2$ impurities. 
The total number of electrons $N$ satisfies $2R \le N \le 4R$ such that there are well-formed local Kondo singlets in the low-energy sector.

At fourth order, perturbation theory is essentially unchanged as compared to the impurity model Eq.\ (\ref{eq:and}), since any local Kondo singlet, consisting of the correlated impurity coupled to an B-sublattice site, is surrounded by uncorrelated A-sublattice sites, and thus the same virtual processes lead to the same effective interaction. 
Therefore, the resulting effective Hamiltonian only involves A-sublattice sites and excess conduction electrons that are not absorbed in a local Kondo singlet.
The hopping term becomes ineffective since the excess conduction electrons are confined between the local Kondo singlets surrounding each A site.
Hence, we are left with a lattice model of A sites, made up by non-local spins $\ff S_{i,\rm bond}$ referring to the bonding orbital around each B site:
\begin{equation}
H_{\rm eff}
=
-
\frac{z^{2} \alpha}{3}
\sum_{i \in B}
\ff S_{i,\rm bond}^{2}
\, ,
\label{eq:eff}
\end{equation}
with $\ff S_{i,\rm bond} = (1/2) \sum_{\sigma\sigma'} b_{i\sigma}^{\dagger} \ff \sigma_{\sigma\sigma'} b_{i\sigma'}$, where $\ff \sigma$ is the vector of Pauli matrices and where the creation operator of the bonding orbital around $i \in B$ is given by $b_{i\sigma}^{\dagger} = \sum_{j \in A}^{n.n.(i)} c^\dagger_{j\sigma} / \sqrt{z}$, i.e., the bonding one-particle orbital is the symmetric superposition of neighboring A-sublattice orbitals:
\begin{equation}
  | {\rm bond}\, i ,\sigma \rangle = \sum_{j \in A}^{n.n.(i)} | j , \sigma \rangle / \sqrt{z} \qquad (i \in B)
  \; .
\end{equation}
The effective spin-only model, Eq.\ (\ref{eq:eff}), is non-trivial as different non-local spins refer to overlapping orbitals and therefore do not commute. 

There is not much known about this model: 
At half-filling, $N = L+R = 3R$, one can rigorously show that a ferromagnetic \cite{ferro} state with fully polarized magnetic moments of the conduction electrons on the A sites is among the ground states. \cite{TSP15}
Exact diagonalization of small systems suggests \cite{TSP15} that the model has a ferromagnetically ordered ground state in the filling range $2R < N < 4R$ (for lower or higher fillings, local Kondo singlets are broken up). 
An {\em inverse} indirect magnetic exchange (IIME) where the magnetic moments of A-site electrons are coupled ferromagnetically via virtual excitations of the local Kondo singlets has been identified as the main physical mechanism. \cite{STP13,TSP14}
For a one-dimensional depleted Anderson lattice, density-matrix renormalization-group calculations have shown \cite{STP13} that the IIME mechanism gradually crosses over to a {\em conventional} (RKKY) \cite{RK54,Kas56,Yos57} indirect magnetic exchange, also favoring ferromagnetism, when varying $V$ from strong to weak hybridization at fixed $U\gg t$.
This crossover and the mutual interplay between RKKY and IIME mechanisms for the magnetic ground-state properties has recently been discussed in Ref.\ \onlinecite{SHP14} in the context of SU(N) models of ultracold Fermi atoms trapped in optical lattices.

The purpose of the present paper is to study the finite-temperature properties of the depleted Anderson lattice, particularly the stability of the ferromagnetic order against thermal fluctuations. 
From the RKKY theory, one can expect $T_{\rm C} \propto J^2 \propto V^4$ for the Curie temperature at weak $V$ and in a parameter regime where the Schrieffer-Wolf transformation \cite{SW66,SN02} applies such that $J=8V^{2}/U$.
On the other hand, for strong $V$, the effective model Eq.\ (\ref{eq:eff}) suggests that $T_{\rm C} \propto \alpha \propto V^{-4}$.
We therefore expect a pronounced maximum of $T_{\rm C}$ at an intermediate $V$.
This optimal $V$ but also the absolute value of $T_{\rm C}$ are interesting from a fundamental theoretical perspective. 
Not only the strong $V$ dependencies but also the fact that the non-interacting ($U=0$) depleted Anderson lattice exhibits a flat band at the Fermi energy \cite{TSP14} promise a comparatively high value for the critical temperature.
Furthermore, the finite-temperature properties are important for the question whether magnetic correlations and magnetic long-range order induced by the IIME can be verified experimentally.
Candidate systems are magnetic nanostructures on non-magnetic surfaces as their geometrical and magnetic properties can be measured, controlled and manipulated to a high degree on a atomic scale. \cite{ES90,HLH06,Wie09,KWCW11,KWC+12}
Likewise, ultracold-atom systems come into question, due to the rapidly improving experimental techniques in this field 
and particularly due to the recent advances to employ fermionic alkaline-earth atoms to efficiently simulate systems with spin and orbital degrees of freedom.
\cite{KBS+10,SBM+11,PHC13,SMO+13,HFS+14,SHH+14,CMP+14}

Our study is based on two different types of mean-field methods: 
To address the strong-$V$ limit, we apply static mean-field theory to the effective spin model Eq.\ (\ref{eq:eff}). 
Since $\ff S_{i,\rm bond}$ is not a rigid spin with $S=1/2$, a fermion mean-field approach must be employed. 
Using this approximation, a rough estimate of the dependence of the Curie temperature on lattice dimension or coordination number and electron density
is obtained. 
Secondly, we apply dynamical mean-field theory (DMFT) \cite{MV89,GKKR96} to the depleted Anderson lattice. 
For a model with a depleted system of correlated sites, the DMFT can expected to yield reliable results since the electron self-energy is much more local as compared to the dense model.
This has been checked for the $D=1$ dimensional model where essentially exact results are available via the density-matrix renormalization group technique. \cite{STP13}
For ground-state properties of local observables as obtained by DMFT even quantitative agreement has been found.

The paper is organized as follows:
The static and dynamical mean-field methods are introduced along with a discussion of the corresponding results in Secs.\ \ref{sec:mf} and \ref{sec:dmft}, respectively. 
The conclusions are summarized in Sec.\ \ref{sec:con}.

\section{Static mean-field theory}
\label{sec:mf}

\subsection{Depleted Anderson lattice}

\begin{figure}[t]
\centerline{\includegraphics[width=0.33\textwidth]{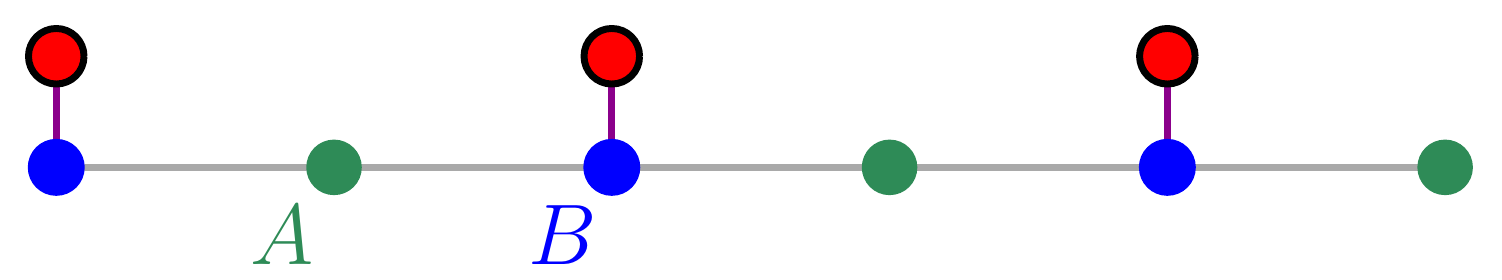} }
\caption{
(Color online) Geometry of the one-dimensional diluted Anderson lattice. 
Red: ``Impurities'' with finite Hubbard interaction. Green and blue: sites of the A and of the B sublattice, respectively.
}
\label{fig:1d}
\end{figure}

\begin{figure}[t]
\centerline{\includegraphics[width=0.4\textwidth]{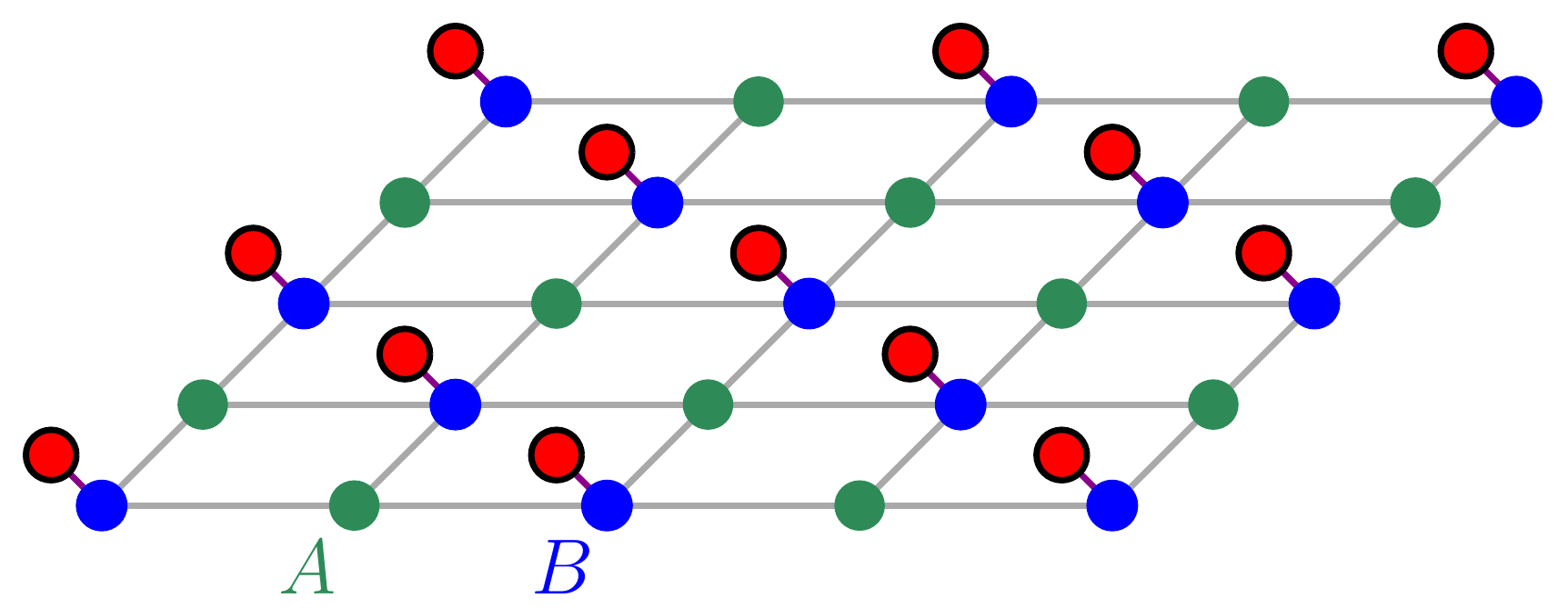} }
\caption{
(Color online) The same as in Figs.\ \ref{fig:1d} and Fig.\ \ref{fig:3d} for the two-dimensional case.
}
\label{fig:2d}
\end{figure}

The Hamiltonian of the depleted Anderson lattice is given by 
\begin{eqnarray}
{\cal H}
&=&
- t \sum_{\langle i,j \rangle ,\sigma} c^\dagger_{i\sigma} c_{j\sigma}
+ V \sum_{i \in B, \sigma} \left(c^\dagger_{i\sigma} f^{\phantom\dagger}_{i \sigma} + \mbox{h.c.} \right)  
\nonumber \\
&+&
U \sum_{i \in B} n_{i \uparrow}^{(f)} n_{i \downarrow}^{(f)} 
- 
\mu \sum_{i, \sigma} n_{i\sigma}^{(c)} 
+ 
(\varepsilon - \mu) \sum_{i \in B,\sigma} n_{i \sigma}^{(f)} 
\, .
\nonumber \\
\label{eq:ham}
\end{eqnarray}
It describes a system of electrons hopping over the sites of a bipartite $D$-dimensional lattice consisting of $L$ sites with periodic boundary conditions.
The two sublattices are denoted by A and B.
We consider a $D=3$ simple-cubic lattice (see Fig.\ \ref{fig:3d}) but also the corresponding one- and two-dimensional cases (Figs.\ \ref{fig:1d} and \ref{fig:2d}).
$c_{i\sigma}^{\dagger}$ creates a conduction electron in a one-particle orbital with spin projection $\sigma=\uparrow, \downarrow$ at the site $i=1,...,L$.
The nearest-neighbor hopping $t=1$ sets the energy scale. 

One-particle orbitals at the B sites of the lattice hybridize with orbitals at $R=L/2$ additional ``impurity'' sites with hybridization strength $V$. 
$f_{i\sigma}^{\dagger}$ creates an electron at the impurity site attached to site $i \in B$ of the sublattice B. 
Furthermore, $n_{i\sigma}^{(c)}=c^\dagger_{i\sigma}c^{\phantom\dagger}_{i\sigma}$ and $n_{i\sigma}^{(f)}=f^\dagger_{i\sigma}f^{\phantom\dagger}_{i\sigma}$ denote the occupation-number operators for A, B and for impurity sites, respectively.
The $f$ orbitals should be considered as magnetic orbitals:
There is a finite repulsive Hubbard interaction $U$ and the one-particle energy is set to $\varepsilon = - U/2$ such that, for strong $U$, the formation of local magnetic moments at the impurity sites is favored. 

The Hamiltonian Eq.\ (\ref{eq:ham}) contains an overall chemical potential $\mu$, i.e., we work with the grand canonical ensemble where $\mu$ is used to fix the average number of particles $\langle N \rangle$.
We will consider the range $2R \le \langle N \rangle \le 4R$ for our calculations. 

Switching off the hopping, i.e.\ $t=0$, defines an atomic limit of the model Eq.\ (\ref{eq:ham}).
The ground state in the atomic limit is highly degenerate. 
For the considered range of the total electron number, each ground state is characterized by completely local Kondo singlets formed on the B and the attached impurity sites binding two electrons per singlet.
The ground state degeneracy is due to the various configurations of remaining electrons on the A sites. 
Their density $n_{\rm A} = \sum_{i \in A, \sigma} \langle n_{i\sigma} \rangle / L_{\rm A}$, where $L_{\rm A}=L/2$ is the number of A sites, can vary within the range $0 \le n_{\rm A} \le 2$.

The depleted Anderson lattice Eq.\ (\ref{eq:ham}) exhibits the conventional U(1) and SU(2) symmetries corresponding to conservation of the total particle number and the total spin. 
For $\mu=0$ the system is half filled, i.e., $\langle N \rangle = 3R$ or $n_{\rm A}=1$, and there is an additional SU(2) isospin symmetry. \cite{TSU97b}
Due to particle-hole symmetry, we can restrict our considerations to the range at and below half-filling.

\subsection{Strong-coupling limit}
 
For strong $V \gg t$, an effective Hamiltonian $\ca H_{\rm eff}$ can be derived by means of fourth-order perturbation theory in $t$ around the degenerate atomic limit. \cite{TSP15}
In this limit the ground state is characterized by local Kondo singlets at the B sites and a residual low-energy dynamics of the A-site electrons which is mediated by virtual high-energy excitations of the local Kondo singlets.
Hence, $\ca H_{\rm eff}$ contains A-site degrees of freedom only. 
There is a very compact and highly symmetric representation of $\ca H_{\rm eff}$ given by Eq.\ (\ref{eq:eff}) with the coupling constant $\alpha$ specified by Eq.\ (\ref{eq:alpha}).
Details of the perturbation theory can be found in Ref.\ \onlinecite{TSP15}.

Here, we rewrite the effective Hamiltonian such that a static mean-field decoupling can be applied in a straightforward way.
To this end, we use the definitions given below Eq.\ (\ref{eq:eff}) to express the non-local spin operators in terms of creators and annihilators for electrons on A sites. 
Furthermore, we switch to a representation in reciprocal space by means of Fourier transformation in the form
\begin{equation}
   c_{i\sigma} = \frac{1}{\sqrt{L_{\rm A}}} \sum_{\ff k \in {\rm BZ}_{A}} e^{i\ff k \ff R_{i}} c_{\ff k\sigma} \qquad (i\in A)
\end{equation}
where $L_{\rm A} = L/2$ and where $\ff k$ is a wave vector in the Brillouin zone BZ$_{A}$ of the reciprocal A sublattice.
Note that the A sublattice is a square lattice for $D=2$ but a b.c.c.\ lattice for the $D=3$ case with a unit cell spanned by the basis vectors $\ff a_{1}, \ff a_{2}, \ff a_{3}$ displayed in Fig.\ \ref{fig:3d}.
With this we get:
\begin{eqnarray}
{\cal H}_{\rm eff}
&=&
\sum_{{\bf k}} (E({\bf k}) - \mu) c_{{\bf k}\sigma}^\dagger c_{{\bf k}\sigma}
\nonumber \\
&+&
\frac{1}{L_A} \sum_{{\bf p}, {\bf q},{\bf k}}
U_{{\bf p}{\bf q}{\bf k}}
c_{{\bf p}\uparrow}^\dagger c_{{\bf p}-{\bf k}\uparrow} c_{{\bf q}\downarrow}^\dagger c_{{\bf q}+{\bf k}\downarrow}  
\: .
\label{eq:heffk}
\end{eqnarray}
The effective one-particle dispersion is given by:
\begin{equation}
  E({\bf k}) = - \frac{D \alpha }{2} \gamma^2({\bf k})
\label{eq:effenergies}
\end{equation}
where $\varepsilon_{0}(\ff k) = -\gamma(\ff k) t$ is the tight-binding dispersion of the $D$-dimensional lattice.
This also determines the $\ff k$ dependence of the interaction parameters of the effective Hamiltonian via:
\begin{equation}
  U_{{\bf p},{\bf q},{\bf k}} = \frac{\alpha}{2} \gamma({\bf p}) \gamma({\bf q}) \gamma({\bf p}-{\bf k}) \gamma({\bf k}+{\bf q}) \: .
\end{equation}
Apparently, the effective Hamiltonian describes itinerant electrons on the A sublattice with an interaction, the $\ff k$ dependence of which corresponds to the non-locality of the quartic parts of the Hamiltonian in real-space representation Eq.\ (\ref{eq:eff}). 

\subsection{Mean-field approximation}
 
Note that in the strong-coupling limit both, the one-particle part as well as the interaction, scale with $\alpha$.
Therefore, the standard mean-field decoupling of the interaction term,
\begin{eqnarray}
c_{{\bf p}\uparrow}^\dagger c_{{\bf p}-{\bf k}\uparrow} c_{{\bf q}\downarrow}^\dagger c_{{\bf k}+{\bf q}\downarrow} 
&\to&
\langle c_{{\bf p}\uparrow}^\dagger c_{{\bf p}-{\bf k}\uparrow} \rangle c_{{\bf q}\downarrow}^\dagger c_{{\bf k}+{\bf q}\downarrow} 
\nonumber \\
&+& 
c_{{\bf p}\uparrow}^\dagger c_{{\bf p}-{\bf k}\uparrow} \langle c_{{\bf q}\downarrow}^\dagger c_{{\bf k}+{\bf q}\downarrow} \rangle
\nonumber \\
&-& 
\langle c_{{\bf p}\uparrow}^\dagger c_{{\bf p}-{\bf k}\uparrow} \rangle \langle c_{{\bf q}\downarrow}^\dagger c_{{\bf k}+{\bf q}\downarrow} \rangle 
\: ,
\label{eq:decouple}
\end{eqnarray}
cannot be controlled by a small parameter but must rather be seen as a Hartree-Fock approach neglecting correlation effects in the low-energy sector and assuming a collinear and homogeneous structure of the magnetic moments.
The formal advantage is that one obtains a mean-field Hamiltonian which allows for a straightforward study of the temperature dependence of the A-site magnetic moment and therewith gives access to the critical (Curie) temperature $T_{\rm C}$. 
However, typical mean-field artifacts must be expected and tolerated.

Using the decoupling (\ref{eq:decouple}) in Eq.\ (\ref{eq:heffk}), we obtain a mean-field Hamiltonian
\begin{equation}
  {\cal H}_{\rm eff} 
  =
  \sum_{{\bf k},\sigma} ({\eta}_\sigma ({\bf k}) - \mu) 
  c_{{\bf k}\sigma}^\dagger c_{{\bf k}\sigma} 
  -
  \alpha
  \frac{L_A}{2}Q_\uparrow Q_\downarrow \, 
\label{eq:mfh}
\end{equation}
which is bilinear in $c^{\dagger}, c$. 
The mean-field dispersion
\begin{equation}
  \eta_\sigma ({\bf k}) = - \frac{\alpha}{2} \left(D-Q_{-\sigma}\right) \gamma^2({\bf k})
\label{eq:mfdis}
\end{equation}
as well as the constant in Eq.\ (\ref{eq:mfh}) depend on the possibly spin-dependent mean field $Q_{\sigma}$ which must be determined self-consistently from the following mean-field equation:
\begin{eqnarray}
  Q_\sigma 
  =
  \frac{1}{L_A} \sum_{\bf k} 
  \gamma^2({\bf k}) \frac{1}{e^{\beta({\eta}_\sigma({\bf k})-\mu)}+1} \: .
\label{eq:qsig}
\end{eqnarray}
Here $\beta=1/T$ and we have chosen units such that $k_B=1$.
The spin-dependent average A-site occupation number, 
$n_{\rm A \sigma} = \sum_{i \in A} \langle n_{i\sigma} \rangle / L_{\rm A}$
is obtained as
\begin{eqnarray}
  n_{\rm A \sigma} = \frac{1}{L_A} \sum_{{\bf k}} \frac{1}{e^{\beta({\eta}_\sigma({\bf k}) - \mu)}+1} \: .
\label{eq:occ}
\end{eqnarray}
With this, the order parameter, i.e., the A-sublattice magnetization, is given by $m_{\rm A} = n_{\rm A \uparrow} - n_{\rm A \downarrow}$.

Numerical calculations are performed by starting with a guess for the chemical potential and solving the coupled system of Eqs.\ (\ref{eq:mfdis}) and (\ref{eq:qsig}) self-consistently for each spin projection. 
From the self-consistent mean field $Q_{\sigma}$, we obtain $n_{\rm A\sigma}$ via Eq.\ (\ref{eq:occ}). 
In an outer self-consistency loop we then adjust the chemical potential until the total filling $n_{\rm A\uparrow} + n_{\rm A\downarrow}$ equals the given filling $n_{\rm A}$. 
In the case of half-filling $n_{\rm A}=1$, calculations are facilitated by particle-hole symmetry which fixes the chemical potential to $\mu=0$.

\subsection{Results}
\label{sec:mfres}

\begin{figure}[t]
\includegraphics[height=0.4\textheight]{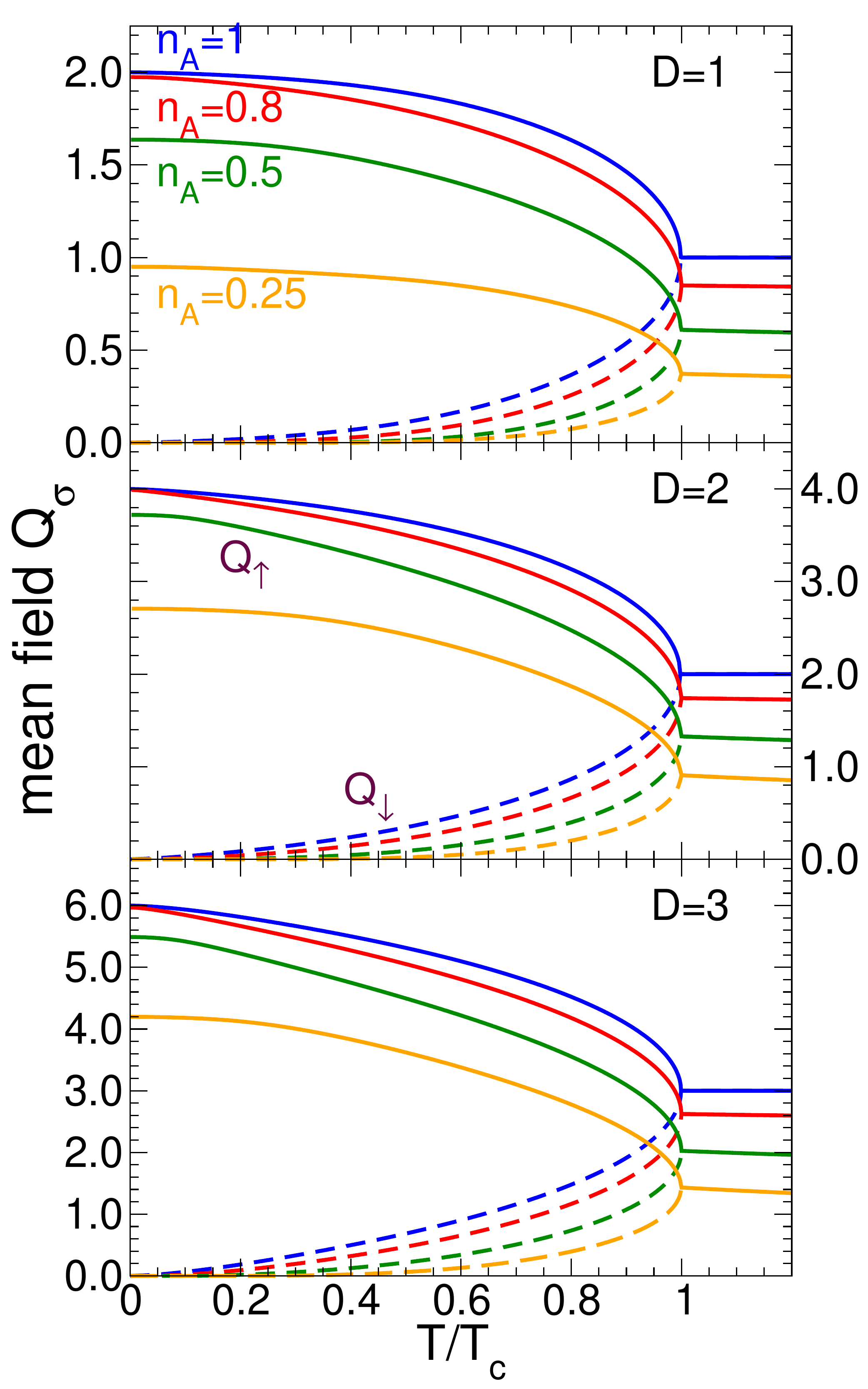}
\caption{(Color online)
Spin-dependent mean field [see Eq.\ (\ref{eq:qsig})] as a function of the reduced temperature $T/T_{\rm C}$ for different dimensions $D$ (top, middle and bottom panel) and different fillings $n_{\rm A}$ as indicated (see top panel) at and below half-filling ($n_{\rm A}=1$). Solid lines: $\sigma=\uparrow$. Dashed lines: $\sigma=\downarrow$.
}
\label{fig:mfq}
\end{figure}
 
\begin{figure}[t]
\includegraphics[height=0.4\textheight]{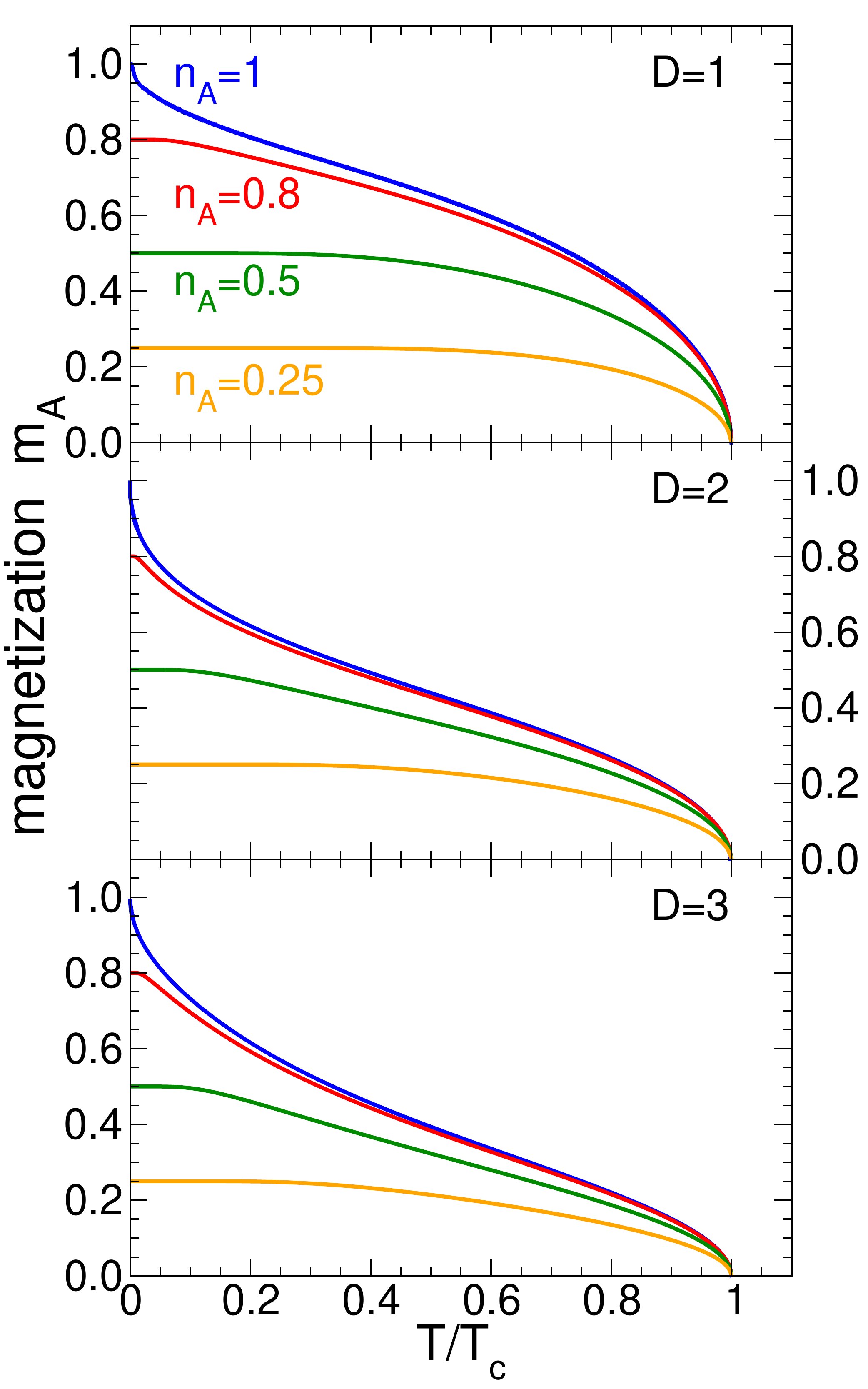} 
\caption{(Color online)
Order parameter $m_{\rm A} = n_{\rm A\uparrow} - n_{\rm A\downarrow}$ [see Eq.\ (\ref{eq:occ})] as a function of $T/T_{\rm C}$ for different dimensions $D$ and fillings $n_{\rm A}$. 
}
\label{fig:mfm}
\end{figure}

Calculations have been performed for lattices with different dimensions $D=1,2,3$ (see Figs.\ \ref{fig:1d}, \ref{fig:2d}, and \ref{fig:3d}, respectively) as well as for different fillings $n_{\rm A}$ at and below half-filling.
Fig.\ \ref{fig:mfq} shows the resulting self-consistent mean fields $Q_{\sigma}$ as functions of the temperature. 
For any $D$ and $n_{\rm A}$, there is a non-zero critical temperature $T_{\rm C}$ below which we find a spontaneous spin-splitting of the mean field. 
This supports the above-mentioned exact-diagonalization results of Ref.\ \onlinecite{TSP15} where a fully polarized magnetic ground state has been found for small one-dimensional systems in the filling range considered.

As can be seen in Fig.\ \ref{fig:mfq} there is only a weak dependence of the mean field on the dimension $D$ -- after rescaling $Q_{\sigma}$ with $D$ or with the coordination number $z$. 
For $T=0$, we have $Q_{\downarrow} = 0$ and thus the $\sigma=\uparrow$ mean-field dispersion simplifies to $\eta_{\uparrow} = - \alpha D \gamma^{2}(\ff k) / 2$ resulting in $Q_{\uparrow} = \mbox{max.}$ and, at half-filling, $Q_{\uparrow} = z$ since particle-hole symmetry enforces $Q_\uparrow - D = D - Q_\downarrow$. 
For $T$ higher than the Curie temperature $T_{\rm C}$, we have $Q_{\uparrow} = Q_{\downarrow}$. 
The spin-independent mean-field is slightly decreasing with increasing $T$, except for half-filling where $Q_{\uparrow} = Q_{\downarrow} = z/2 = \mbox{const.}$ above $T_{\rm C}$. 

Fig.\ \ref{fig:mfm} shows the temperature-dependent magnetization for the different fillings and dimensions. 
At zero temperature, the system is always fully polarized, i.e., $n_{\rm A\downarrow} = 0$ and $n_{\rm A\uparrow} = n_{\rm A}$.
Similar to the discussion of the mean fields, after rescaling the temperature with the respective Curie temperature $T_{\rm C}$, there is a weak dependence of $m_{\rm A}$ on the dimension $D$ at finite $T$. 
The phase transition from the ferro- to the paramagnetic state at $T_{\rm C}$ is of second order for any $n_{\rm A}$.
Close to the Curie point, we find a critical behavior of $m_{\rm A}$ characterized by the (mean-field) critical exponent for the magnetization $\beta=0.5$, as expected.

The only unexpected result consists in the unconventional $T$-dependence of $m_{\rm A}$ at half-filling.
While at low temperatures the missing feedback of long-wavelength spin excitations explains the absence of a power-law $T$ dependence, one would expect, as a typical mean-field behavior, an exponential convergence of $m_{\rm A}(T)$ for $T\to 0$ with a negative curvature and a vanishing slope
$\lim_{T \to 0} d m_{\rm A} / dT =0$.
However, for $n_{\rm A}=1$, Fig.\ \ref{fig:mfm} shows an inflection point of $m_{\rm A}(T)$ at a finite temperature, which is increasing with increasing $D$, and an unusual upturn of $m_{\rm A}$ for $T\to 0$. 
Closer inspection of the data shows that the slope is diverging:
\begin{equation}
  \frac{d m_{\rm A}}{dT} \sim - \frac{1}{\sqrt{\alpha D T}} \rightarrow -\infty  \quad\quad\quad (D=1,D=3) 
\label{eq:m1}
\end{equation}
and 
\begin{equation}
  \frac{d m_{\rm A}}{dT} \sim \frac{\ln( T/\alpha) }{\sqrt{\alpha D T}}  \rightarrow -\infty   \quad\quad (D=2) \, .
\label{eq:m2}
\end{equation}
The reason of this behavior is a van Hove singularity of the spin-dependent mean-field local density of states at the Fermi edge and is discussed in Appendix \ref{sec:app}.

\begin{figure}[t]
\centerline{\includegraphics[width=0.4\textwidth]{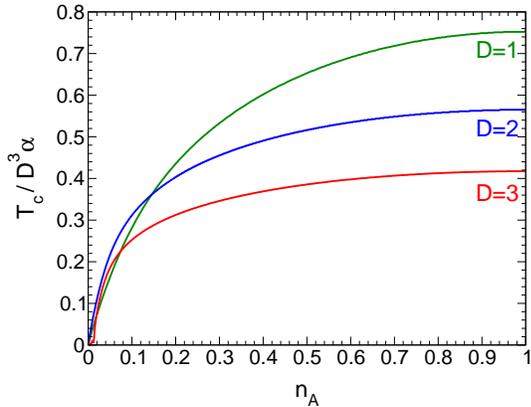}}
\caption{(Color online)
Filling dependence of the Curie temperature for lattices with different dimensions as obtained from the static mean-field theory. 
Note that $T_{\rm C}$ is rescaled by $D^{3}$ and given in units of the coupling constant $\alpha$.
}
\label{fig:tc}
\end{figure}

From the temperature-dependence of the order parameter we can read off the Curie temperature. 
This is plotted in Fig.\ \ref{fig:tc} for different $D$ as functions of the filling $n_{\rm A}$.
Since $T_{\rm C}(n_{\rm A}) = T_{\rm C}(2-n_{\rm A})$ due to particle-hole symmetry, we restrict ourselves to the range $n_{\rm A}\le 1$.
Clearly, the Curie temperature must be proportional to the coupling constant $\alpha$ as there is a single energy scale in the effective Hamiltonian Eq.\ (\ref{eq:eff}).

Its geometry dependence is more interesting:
Namely, $T_{\rm C}$ is by no means proportional to the coordination number as it typical for many mean-field approaches but is much more rapidly increasing with increasing $z$ (note that the numerical results are scaled by a factor $D^{3}$ in Fig.\ \ref{fig:tc}).
This finding is not related to singularities in the density of states as it holds for any filling. 
We attribute the unconventionally high $T_{\rm C}$ to the non-locality of the interaction in the effective Hamiltonian and to the resulting $\ff k$-dependent contribution of the mean field to the mean-field dispersion in Eq.\ (\ref{eq:mfdis}).

\section{Dynamical mean-field theory}
\label{sec:dmft}

\subsection{General theory}

Dynamical mean-field theory (DMFT) \cite{MV89,GKKR96} neglects the feedback of non-local, e.g.\ magnetic correlations, on the local self-energy and the local one-particle Green's function but correctly accounts for all local correlations. 
This represents a decisive step beyond the static mean-field approach.
Particularly, the DMFT is able to describe the formation of local magnetic moments already in the paramagnetic phase of a lattice model of itinerant electrons, such that the phase transition between the paramagnetic and the ferromagnetic phase at $T_{\rm C}$ can be understood as a transition between well-formed but disordered moments and long-range order.
This is opposed to the static theory where the local moments essentially vanish above the Curie point.

It is important to note that the feedback of non-local correlations neglected within single-site DMFT is much weaker for the depleted Anderson lattice considered here as compared to a lattice fermion model with a dense system of correlated sites.
This can be understood in the following way:
Formally, the only approximation to be tolerated within DMFT is the locality of the self-energy. 
For a dense lattice model, such as the Hubbard model, for example, the DMFT becomes exact in the limit of infinite spatial dimensions $D \to \infty$ since the nearest-neighbor elements of the self-energy scale as $1/D^{3/2}$ as can be inferred from its diagram expansion. \cite{MH89c}
This is related to the scaling $1/D^{\| i - j \| /2}$ of the bare propagator, where $d \equiv \| i - j \|$ is the Manhattan distance between the orbitals at sites $i$ and $j$ of a hyper-cubic lattice of dimension $D$.
For a depleted Anderson lattice on a high-dimensional bipartite lattice with a Manhattan distance $d$ between the correlated sites, this also implies that the non-local elements of the self-energy exponentially diminish with increasing $d$. 
A completely local self-energy is realized in the single-impurity limit $d \to \infty$.
For the Hubbard model ($d=1$) and the periodic Anderson model ($d=3$) and for low dimensions, quantitative studies have been performed within second-order perturbation theory. \cite{SC90,SC91,PN97c}
The case studied here corresponds to $d=4$ but there are two, possibly largely different hopping parameters, $t$ and $V$.
For the ground state of the depleted Anderson lattice (with $d=4$) in $D=1$ dimension, a direct comparison between DMFT and essentially exact results obtained by the density-matrix renormalization group (DMRG) method has been performed in Ref.\ \onlinecite{STP13}, and excellent agreement has been found for static local observables in the entire $V/t$ regime.
Comparing with DMRG, a quantitative discussion of the artifacts of the DMFT has been given in Ref.\ \onlinecite{TSRP12} for a $D=1$ tight-binding model with two Anderson impurities.
Concluding, we therefore expect that the DMFT yields reliable results.

DMFT is easily adapted to the model Eq.\ (\ref{eq:ham}):
For any dimension $D$, there are three sites in a primitive unit cell of the lattice (see Figs.\ \ref{fig:3d}, \ref{fig:1d} and \ref{fig:2d}). 
Hence, the single-particle Green's function $\ff G_{\ff k}(\omega)$ is a $3\times 3$-matrix for any wave vector $\ff k$ in the first Brillouin zone of the A sublattice and for any one-particle excitation energy $\omega$. 
Summation over $\ff k$ provides us with the local Green's function with, say, the (3,3) element referring to the impurity Green's function 
$\langle \langle f_{i\sigma} ; f^{\dagger}_{i\sigma} \rangle \rangle_\omega$. 
Using Dyson's equation, this can be obtained from the local self-energy $\Sigma(\omega)$ as 
\begin{equation}
 G_{\rm loc}^{(\alpha\beta)}(\omega) =
 \frac{2}{L} \sum_{\ff k \in {\rm BZ}_{A}}
 \left[ 
 \frac{ 1 }{ \omega + \mu - \ff \varepsilon(\ff k) - \ff \Sigma(\omega) }
 \right]_{\alpha\beta}
 \; .
\label{eq:green}
\end{equation}
Here, $\alpha, \beta = 1,2,3$ label the different sites in a unit cell. 
Furthermore, $\ff \Sigma(\omega)$ is a $3\times 3$ diagonal matrix with $\Sigma_{33}(\omega)=\Sigma(\omega)$ and $\Sigma_{11}(\omega)=\Sigma_{22}(\omega)=0$, and 
\begin{equation}
\ff \varepsilon(\ff k)
=
\left(
\begin{array}{ccc}
0 & \varepsilon_{0}(\ff k) & 0 \\
\varepsilon_{0}(\ff k) & 0 & V \\
0 & V & \varepsilon \\
\end{array}
\right)
\end{equation}
is the lattice Fourier transform of the hopping parameters with $\varepsilon_{0}(\ff k) = -\gamma(\ff k) t$.

The DMFT self-energy $\Sigma(\omega)$ is obtained as the impurity self-energy of an effective Anderson impurity model specified by the Hubbard-$U$ and a hybridization function that is fixed by the self-consistency equation of DMFT as
\begin{equation}
  \Delta(\omega) = \omega + \mu - \varepsilon - \Sigma(\omega) - \frac{ 1 }{ G_{\rm loc}^{(33)}(\omega)} \: .
  \label{eq:hyb}
\end{equation}
Here, the impurity one-particle energy is given by $\varepsilon$, and $\Sigma(\omega)$ must be determined self-consistently with Eq.\ (\ref{eq:green}). 

To compute the self-energy of the effective impurity problem at finite temperature $T$, we employ the continuous-time quantum Monte-Carlo method \cite{RSL05,GML+11} and the hybridization expansion of the action of the effective impurity model. \cite{WCdM+06}
Configurations are sampled by means of the Metropolis-Hastings algorithm. \cite{MRR+53,Has70}
As the Hubbard interaction is of density-density type, we can use the highly efficient segment-picture variant of the approach and, following Ref.\ \onlinecite{HPW12}, directly measure the impurity self-energy $\Sigma_{\sigma}(i \omega_n)$ on the fermionic Matsubara frequencies $i\omega_{n}$.

\subsection{Results}

From the results of static mean-field theory for the effective low-energy model Eq.\ (\ref{eq:eff}) in the strong $V$ limit (see Fig.\ \ref{fig:tc}) we infer that the Curie temperature is at a maximum for half-filling. 
More generally, we expect that at half-filling the stability of a ferromagnetically ordered state against thermal fluctuations is the highest not only for strong $V$ but also for weak $V$, where the period of the RKKY interaction is commensurate with the positions of the correlated sites on the lattice.
Furthermore, at half-filing and for the considered lattice geometries, the RKKY interaction is ferromagnetic.
We will therefore restrict ourselves to the particle-hole symmetric case with the chemical potential fixed at $\mu=0$ and with the one-particle energy of the impurities set to $\varepsilon = - U/2$ [see Eq.\ \ref{eq:ham})].
We also fix the Hubbard interaction at an intermediate value $U=8$ for the rest of the paper. 
To discuss the crossover from the RKKY limit to the regime of the inverse indirect magnetic exchange, we consider different hybridization strengths $V$.

By carrying out a sum over Matsubara frequencies, one may easily compute the average spin-dependent occupation numbers on the A and B sites, $n_{\rm A\sigma} = \langle c^{\dagger}_{A\sigma} c_{\rm A\sigma} \rangle$ and $n_{\rm B\sigma} = \langle c^{\dagger}_{B\sigma} c_{\rm B\sigma} \rangle$, from the local Green's function Eq.\ (\ref{eq:green}), once self-consistency has been achieved.
The average occupation numbers of the impurity site, $n_{\rm imp, \sigma} = \langle f^{\dagger}_{\sigma} f_{\sigma} \rangle$, can be obtained in the same way or, equivalently, can be measured within CT-QMC directly. 

\begin{figure}[t]
\centerline{\includegraphics[width=0.5\textwidth]{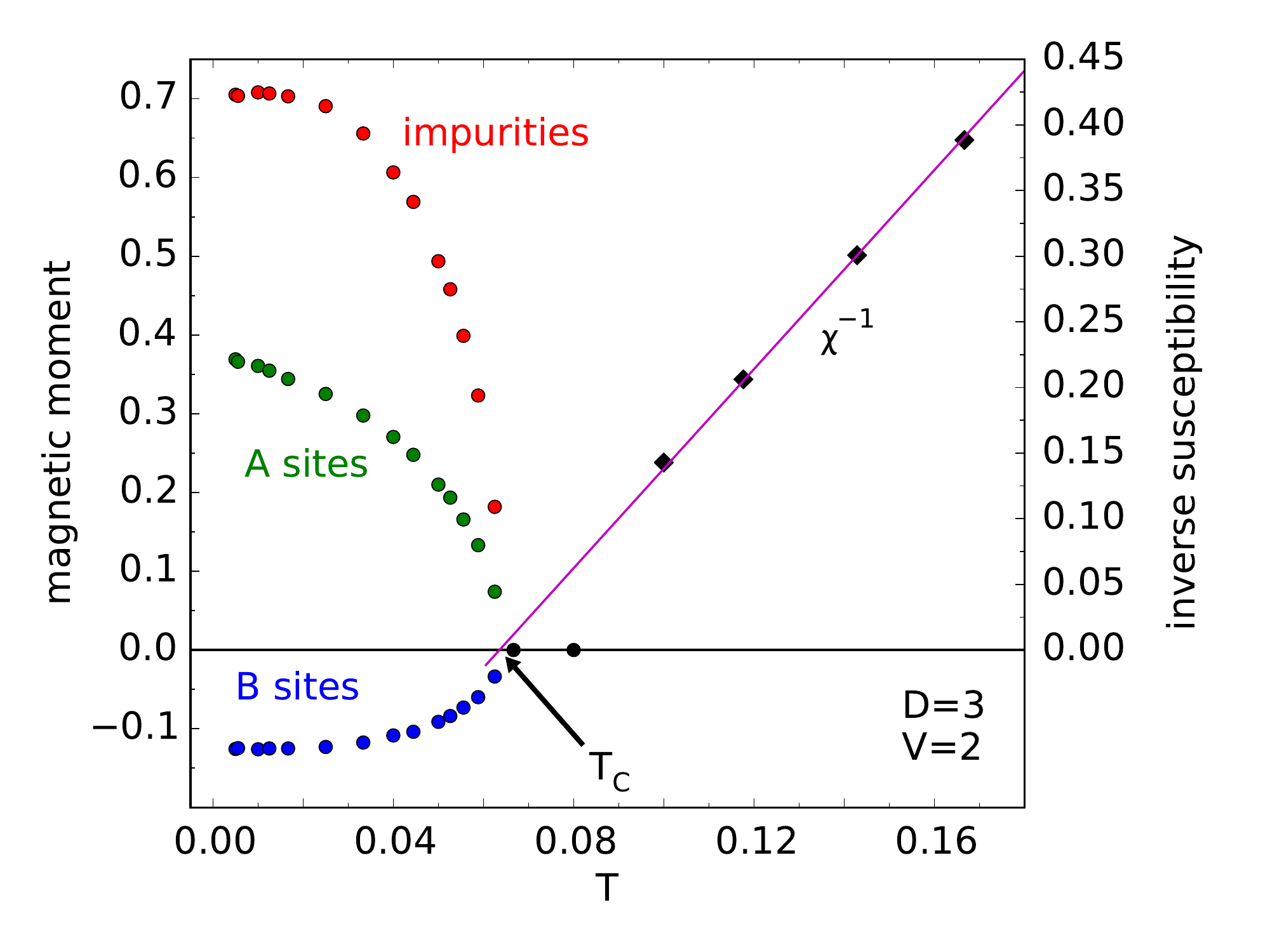} }
\caption{(Color online) 
Ordered magnetic moments $m_{\rm A}$, $m_{\rm B}$, $m_{\rm imp}$ (circles) on the A sites, the B sites and the impurity sites, respectively, and the inverse homogeneous static impurity magnetic susceptibility $\chi^{-1}$ (diamonds) as functions of temperature $T$ as obtained by DMFT for the $D=3$-dimensional depleted Anderson lattice (see Fig.\ \ref{fig:3d}).  
Hubbard interaction: $U=8$, hybridization strength: $V=2$. 
The line indicates a linear fit to the trend of $\chi^{-1}(T)$. 
The temperature and energy scales are fixed by the nearest-neighbor hopping $t=1$ [see Eq.\ (\ref{eq:ham})].
}
\label{fig:allD3V2}
\end{figure}

\begin{figure}[b]
\centerline{\includegraphics[width=0.5\textwidth]{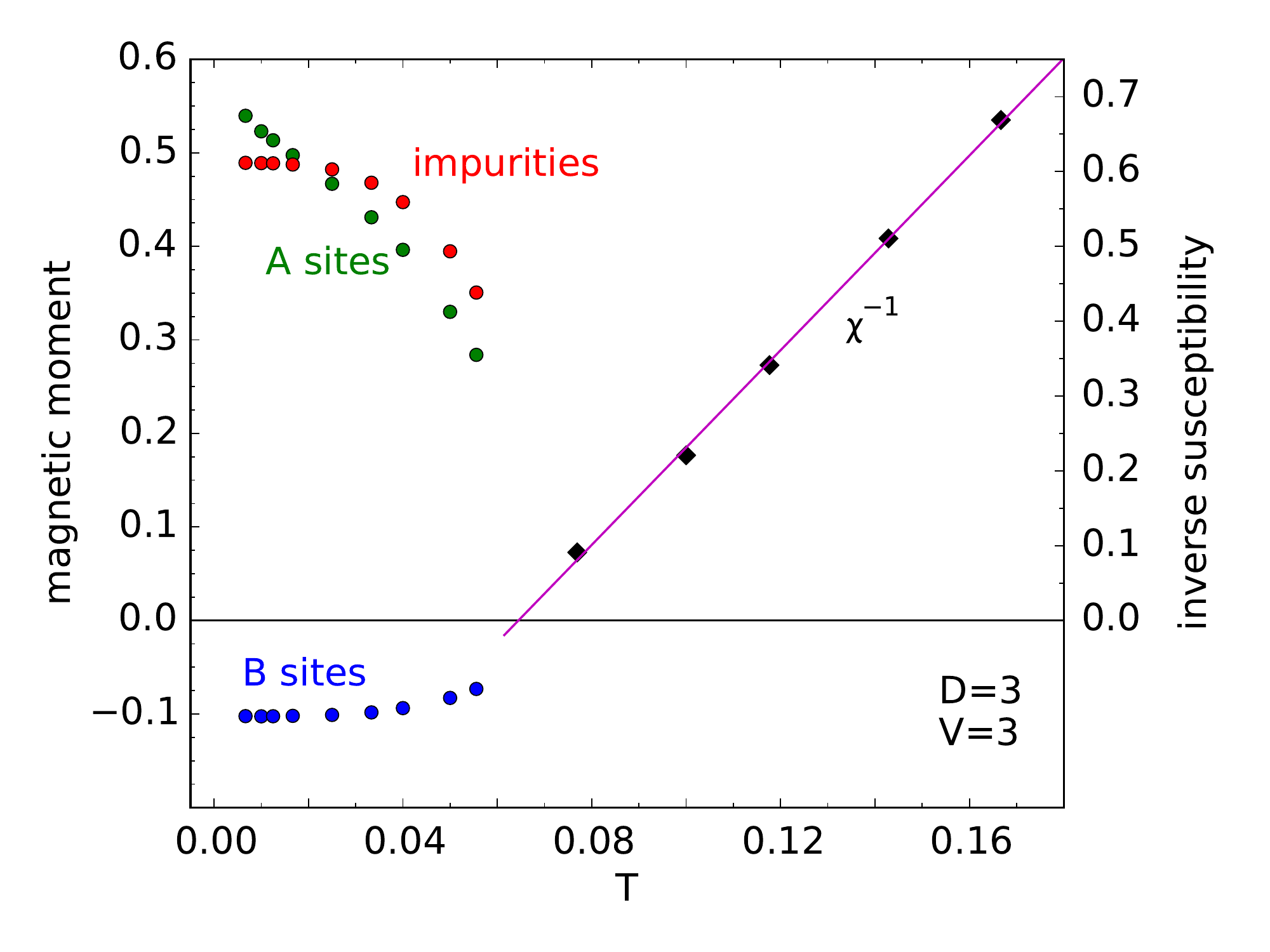} }
\caption{(Color online) 
The same as in Fig.\ \ref{fig:allD3V2} but for $V=3$.
}
\label{fig:allD3V3}
\end{figure}

We explicitly allow for symmetry-broken states with finite ordered magnetic moments $m_{\rm A}$, $m_{\rm B}$ and $m_{\rm imp}$, where we have defined $m_{\rm A} = n_{\rm A\uparrow} - n_{\rm A\downarrow}$, and $m_{\rm B}$, $m_{\rm imp}$ analogously. 
It is found that magnetic solutions of the DMFT equations are easily stabilized in the entire range of hybridization strengths $V$ considered (but for sufficiently low temperatures) by starting the self-consistency cycle with a slightly spin-asymmetric initial self-energy. 
In addition, we also compute the homogeneous static impurity spin susceptibility of the paramagnetic state $\chi = \partial m_{\rm imp} / \partial B |_{B=0}$.
Here, $B$ is the strength of a homogeneous magnetic field coupling to the $z$-component of the total impurity spin as $H \mapsto H - B \sum_{i \in B} (f^{\dagger}_{i\uparrow} f_{i\uparrow} - f^{\dagger}_{i\downarrow} f_{i\downarrow})$ where $H$ is given by Eq.\ (\ref{eq:ham}).

Fig.\ \ref{fig:allD3V2} shows the results of a DMFT calculation at $V=2$ for the $D=3$-dimensional lattice with $L=52^{3}$ sites, with additional $R=L/2$ impurites and periodic boundary conditions (see Fig.\ \ref{fig:3d}). 
This is fully sufficient to ensure that the results do not significantly depend on $L$.
Statistical errors of the quantities shown in this and in the following figures are smaller than the size of the symbols.
A typical Monte-Carlo run consists of more than $10^{7}$ sweeps, and each sweep of more than $k$ Monte-Carlo steps with $k$ being the average expansion order. 
Less than 50 DMFT iterations are sufficient for convergence of the results within the statistical error.

For high temperatures the system is in a paramagnetic state.
The inverse susceptibility $\chi^{-1}$ shows a linear Curie-Weiss trend from which one can safely estimate the value for the Curie temperature $T_{\rm C} \approx 0.064$. 
$\chi$ is calculated from the magnetic moments induced by an explicitly applied homogeneous field for sufficiently weak field strengths in the linear-response regime (typically $B<0.01$).

The transition to the ferromagnetic state at low temperatures appears to be of second order, and
the data for the ordered magnetic moments are consistent with a linear temperature trend of $m^{2}$ close to $T_{\rm C}$, i.e.\ $m^{2} \propto (T_{\rm C}  - T)$. 
This implies a critical exponent $\beta=0.5$ as it must be expected for a DMFT calculation. 
Note, however, that due to critical slowing down, it becomes progressively more difficult to stabilize symmetry-broken DMFT solutions for temperatures close to $T_{\rm C}$.
The double occupancy at the impurity site $d_{\rm imp} = \langle n_{\rm imp, \uparrow} n_{\rm imp, \downarrow} \rangle$, and thus the local magnetic moment $\ff S_{\rm imp}^{2} = 3 ( 1- 2 d_{\rm imp}) / 4$ turns out to be almost constant in the entire temperature range considered: $d_{\rm imp} \approx 0.077$. 
In particular, the moment does not change significantly across the phase transition.
The transition to the symmetry-broken state must therefore be characterized as long-range ordering of local magnetic moments that are preformed at higher temperatures. 
This is a typical effect of strong correlations and opposed to simple Hartree-Fock-like (or Stoner-like) phase transitions where the local magnetic moment forms right at $T_{\rm C}$.

\begin{figure}[t]
\centerline{\includegraphics[width=0.5\textwidth]{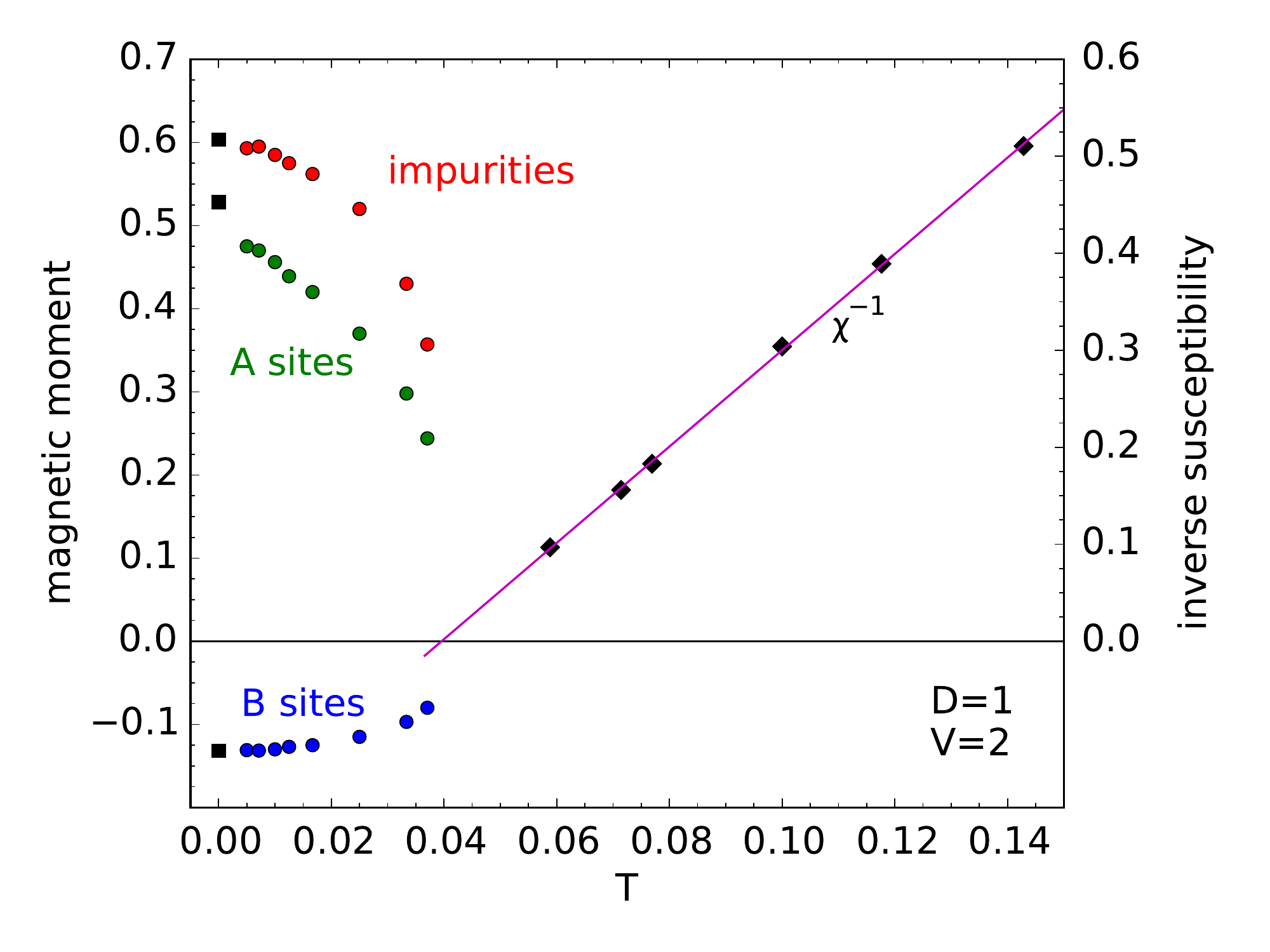}} 
\caption{(Color online) 
$m_{\rm A}$, $m_{\rm B}$, $m_{\rm imp}$ (circles) and $\chi^{-1}$ (diamonds) as functions of $T$, as in Fig.\ \ref{fig:allD3V2} but for the $D=1$-dimensional depleted Anderson lattice (see Fig.\ \ref{fig:1d}).  
Hubbard interaction: $U=8$, hybridization strength: $V=2$. 
DMRG data for $T=0$ (squares) from Ref.\ \onlinecite{STP13} are included for comparison.
}
\label{fig:allD1V2}
\end{figure}

\begin{figure}[t]
\centerline{\includegraphics[width=0.5\textwidth]{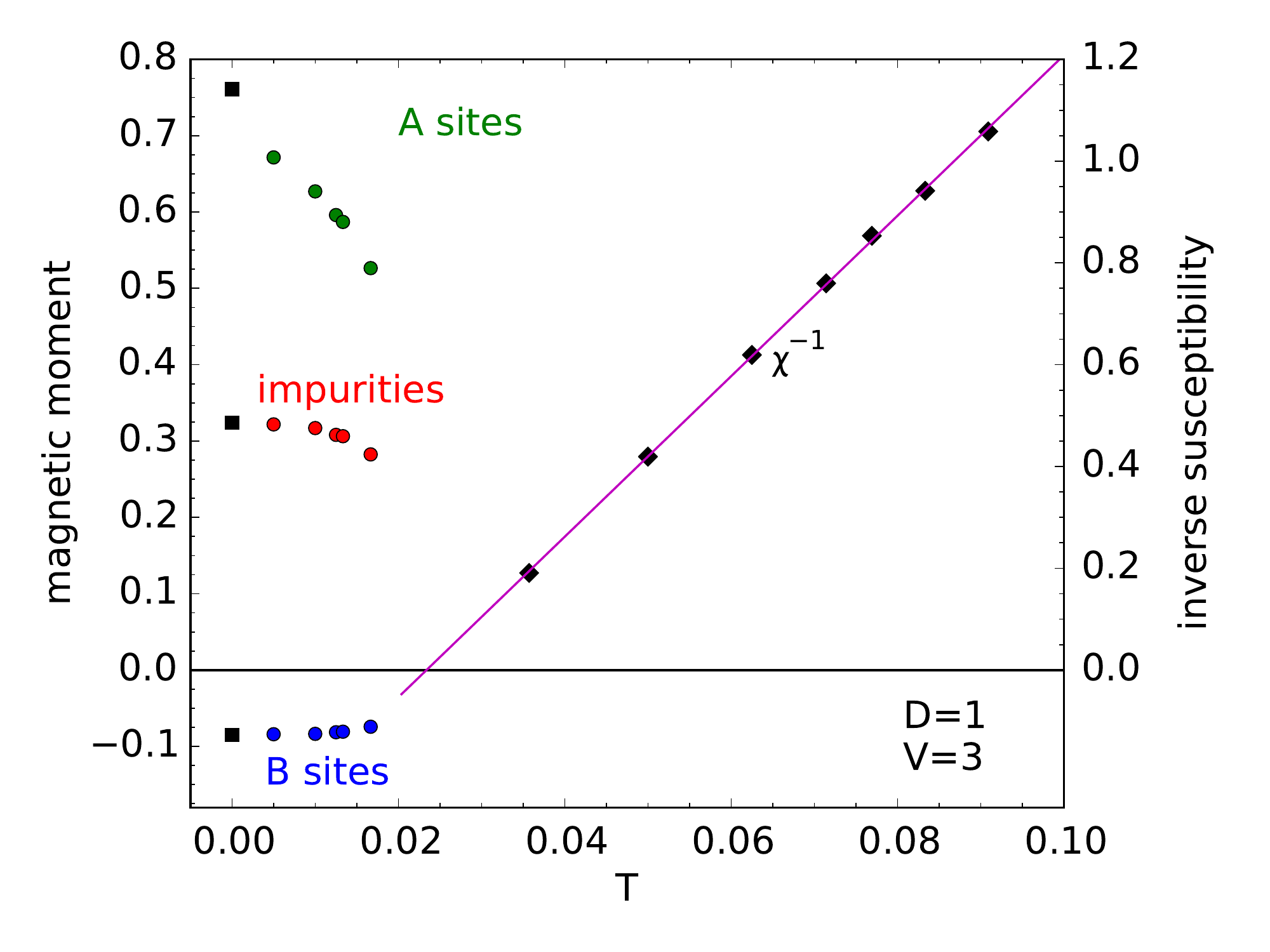}}
\caption{(Color online) 
The same as in Fig.\ \ref{fig:allD1V2} but for $V=3$.
}
\label{fig:allD1V3}
\end{figure}

The low-temperature state of the system actually displays ferri-magnetic order since the magnetic moment at the B sites is antiferromagnetically aligned ($m_{\rm B}<0$) to the moments at the impurities and the A sites ($m_{\rm imp}, m_{\rm A}>0$).
This alignment is reminiscent of the antiferromagnetic coupling in the Kondo limit of the model, i.e.\ for $V\to 0$, where an antiferromagnetic effective exchange interaction (Kondo coupling) of strength $J=8V^{2}/U$ emerges between B sites and impurities in the low-energy sector. \cite{SW66,SN02} 
In the weak-coupling limit $V\to 0$, one furthermore expects that well-formed local magnetic moments appear at the impurity sites since charge fluctuations are strongly suppressed.
Ferromagnetic coupling of these moments via the RKKY exchange then implies $|m_{\rm imp}| \to 1$, while $m_{\rm A}, m_{\rm B} \to 0$. 
For $V=2$, we are still in the RKKY regime since the A-site moment is clearly smaller than the moment on an impurity site.

As Fig.\ \ref{fig:allD3V3} demonstrates, however, this changes with increasing $V$. 
For $V=3$, we find $m_{\rm A} > m_{\rm imp}$ at low temperatures indicating the crossover from the RKKY regime to the strong-$V$ limit. 
In the strong-coupling limit $V \gg t$, almost localized ``Anderson singlets'' are formed by the magnetic moments at B and impurity sites, and thus $m_{\rm B}, m_{\rm imp} \to 0$. 
The presence of local singlets at the B sites implies that electrons on the remaining A sites are very efficiently localized such that well-formed local moments emerge. 
Those moments couple ferromagnetically via the inverse indirect magnetic exchange, \cite{STP13,TSP14,TSP15} i.e.\ by virtual excitations of the Anderson singlets, and thus $m_{\rm A} \to 1$. 
This picture well explains that $m_{\rm A} > m_{\rm imp}$ in Fig.\ \ref{fig:allD3V3}.

It is instructive to compare the results for the $D=3$ lattice with those obtained for $D=1$ (see Fig.\ \ref{fig:1d}).
Figs.\ \ref{fig:allD1V2} and \ref{fig:allD1V3} show results for the ordered magnetic moments and the impurity magnetic susceptibility for a chain geometry with $L=50$ sites with periodic boundary conditions. 
The overall trends seen in the figures are similar to those found for $D=3$ but the crossover from the RKKY to the IIME regime appears at lower hybridization strength $V$ as can be inferred from the fact that $m_{\rm A}$ is considerably higher than $m_{\rm imp}$ already for $V=3$. 

Furthermore, the Curie temperature is seen to {\em decrease} with increasing $V$ in this regime; $T_{\rm C}$ drops by about a factor two when increasing the hybridization strength from $V=2$ to $V=3$. 
This can consistently be explained by referring to the strong-$V$ limit where the effective model Eq.\ (\ref{eq:eff}) applies and where the only energy scale is given by the coupling $\alpha$ [see Eq.\ (\ref{eq:alpha})] which decreases with increasing $V$.

This also means that the crossover regime shifts to stronger hybridizations strengths with increasing lattice dimension or coordination number.
This must be kept in mind when comparing $T_{\rm C}$ obtained for different dimensions $D$ at constant $V$.
At $V=2$, the Curie temperature does not depend very much on $D$:
We find $T_{\rm C}=0.040$ for $D=1$, $T_{\rm C}=0.059$ for $D=2$ 
and $T_{\rm C}=0.064$ for $D=3$. 
This is easily explained as a balance between two counteracting effects, namely an increase of $T_{\rm C}$ with increasing $D$ characteristic for a mean-field theory on the one hand and the mentioned shift of the crossover regime resulting in a lower $T_{\rm C}$ on the other hand.


\begin{figure}[t]
\centerline{\includegraphics[width=0.4\textwidth]{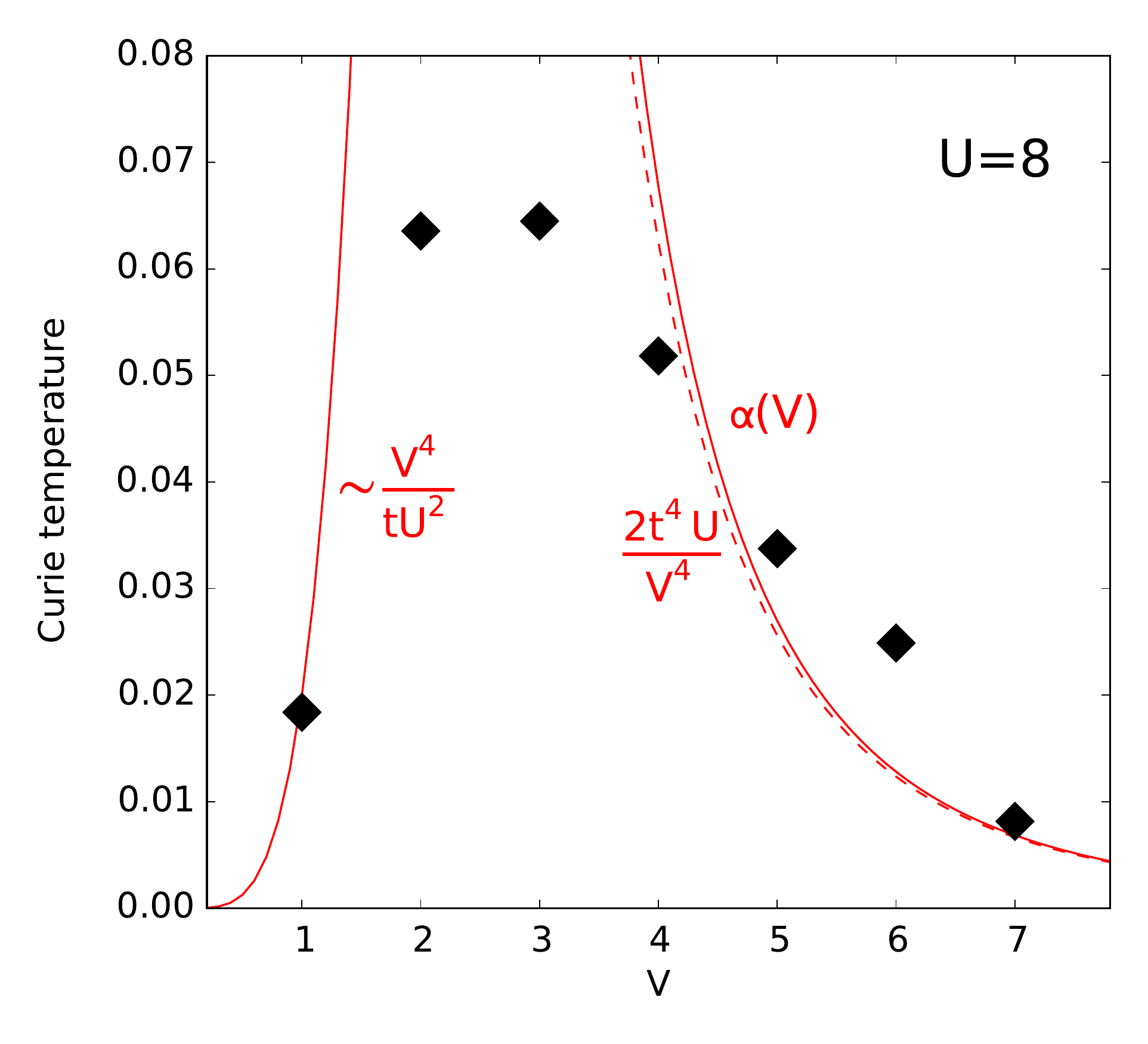} }
\caption{(Color online) 
Curie temperature $T_{\rm C}$ for the $D=3$-dimensional depleted Anderson lattice at $U=8$ and half-filling as a function of the hybridization strength $V$.
Points are obtained via $\chi^{-1}(T_{\rm C})=0$ by extrapolating the linear temperature trend of the inverse susceptibility $\chi^{-1}(T)$.
Solid lines: a dependence of $T_{\rm C}(V) \propto V^4 / tU^{2}$ is expected for $V \to 0$. 
For strong $V$, the data are consistent with $T_{\rm C}(V) \propto \alpha(V)$. 
Dashed line: $T_{\rm C}(V) \propto 2 t^{4} U / V^{4}$ represents a good approximation to $\alpha(V)$ at $U=8$.
}
\label{fig:tcofv}
\end{figure}

Obviously, the $D=1$ and $D=2$ results are not consistent with the Mermin-Wagner theorem \cite{MW66} which excludes spontaneous breaking of the SU(2) spin rotation symmetry for $D\le 2$ at finite temperatures.
As a matter of course, it {\em cannot} be satisfied within a static or within dynamical mean-field theory since long wave-length magnetic excitations do not feed back to the single-particle self-energy.
We nevertheless expect that the finite $T_{\rm C}$ predicted by DMFT is physically significant even for $D=1$ (and $D=2$) and indicates the onset of ferromagnetic ordering of the magnetic moments on {\em intermediate} length scales. \cite{VMS+11}
This corresponds to a thermodynamically stable ferromagnet only if the SU(2) symmetry is broken explicitly, e.g., due to the presence of additional anisotropic terms in the Hamiltonian.

For $D=1$ and in the low-temperature limit the DMFT agrees well with essentially exact data obtained by means of the density-matrix renormalization group (DMRG) method \cite{STP13} at zero temperature.
The extrapolation of the DMFT results for the ordered magnetic moments $m_{\rm imp}$ and $m_{\rm B}$ to $T=0$ perfectly matches with the DMRG data, see black squares in Figs.\ \ref{fig:allD1V2} and \ref{fig:allD1V3}.
As concerns the magnetic moment on the A sites, we expect the same unconventional $T$-dependence that has been discussed in the context of static mean-field theory in Sec.\ \ref{sec:mfres}, i.e.\ an upturn of $m_{\rm A}$ for $T\to 0$, consistent with the $T=0$ DMRG data, which is induced by the van Hove singularity of the spin-dependent local density of states at the Fermi edge.

For the $D=3$ lattice, we have systematically computed $T_{\rm C}$ as a function of $V$ at fixed $U=8$. 
Results as obtained from by linear fits to the temperature trend of the $\chi^{-1}(T)$ are shown in Fig.\ \ref{fig:tcofv}. 

In the weak-coupling limit $V \to 0$, the Curie temperature is expected to be solely determined by the strength of the RKKY interaction and thus to scale as $T_{\rm C} \propto J_{\rm RKKY} \propto J^{2} \propto V^{4}$ with $V$. 
For the strong-coupling or IIME limit, the only energy scale of the effective low-energy theory is given by $\alpha$ and thus $T_{\rm C} \propto \alpha$ (see solid lines in Fig.\ \ref{fig:tcofv}).
For $U=8$, a good approximation is $T_{\rm C} \propto V^{-4}$, see Eq.\ (\ref{eq:alpha}) and the dashed line in Fig.\ \ref{fig:tcofv}.

The Curie temperature is at its maximum $T_{\rm C, opt} \approx 0.07$ for a hybridization strength of about $V_{\rm opt} \approx 2.5$. 
$T_{\rm C, opt}$ is almost an order of magnitude smaller than the maximum N\'eel temperature of the $D=3$ Hubbard model at half-filling \cite{KJMP05} and also an order magnitude smaller than typical Curie temperatures of the Hubbard model with asymmetric free density of states, as obtained for lower fillings by DMFT. \cite{Ulm98}
The same holds if compared with DMFT estimates for the Curie temperature of the standard periodic Anderson model. \cite{MN00} 

\section{Summary}
\label{sec:con}

The present study has demonstrated that the Anderson-lattice model with a regularly depleted system of localized orbitals at every second site supports ferromagnetic long-range order which exhibits, depending on the hybridization strength $V$, a high stability against thermal fluctuations.
The temperature-dependent magnetism has been investigated systematically for different coupling strengths and electron densities.

The depleted Anderson lattice model has been considered beforehand
to study fundamental questions of magnetic coupling mechanisms \cite{STP13,TSP14} 
and to describe artificial Kondo systems realized as ultracold atoms trapped in optical lattices. \cite{SHP14}
It is related to two-dimensional superlattices consisting of periodic arrangements of $f$-electron- and non-interacting layers \cite{PTK13}
and may be used to describe systems of magnetic atoms on non-magnetic metallic surfaces where a manipulation of the adatom geometry and a precise mapping of magnetic couplings is accessible to scanning-tunneling techniques on an atomic scale.
\cite{ES90,HLH06,Wie09,KWCW11,KWC+12}

We have employed two different types of mean-field approaches: 
(i) static mean-field theory of the effective low-energy model that emerges at strong couplings $V$ within fourth-order perturbation theory, and (ii) dynamical mean-field theory of the full model using continuous-time quantum Monte-Carlo as impurity solver. 
The Curie temperature is obtained by computing the temperature dependence of the magnetic moments as well as by the divergence of the homogeneous static magnetic susceptibility. 
The maximal $T_{\rm C}$ is found at half-filling and for intermediate hybridization strengths:

For weak $V$, magnetic order is induced by the standard effective RKKY interaction between the local magnetic moments formed at the correlated impurity sites. 
For the geometry considered and at half-filling, the RKKY interaction is ferromagnetic. 
The Curie temperature scales with $V^{4}$ in this limit.
For strong $V$, on the other hand, the recently proposed inverse indirect magnetic exchange also leads to ferromagnetic order. 
In this limit the impurity magnetic moments are Kondo screened and form almost local Kondo singlets on a high-energy scale $V$ which localize the fraction of conduction electrons not taking part in the screening. 
Those conduction electrons develop local magnetic moments which are ferromagnetically coupled by virtual excitations of the local Kondo singlets on an energy scale $\alpha$ [see Eq.\ (\ref{eq:alpha})]. 
Therefore, $T_{\rm C}$ scales with $\alpha \sim V^{-4}$ for fixed $U$ in this limit.

While the numerical data obtained for different $V$ appear to be consistent with the expected trends, it turned out to be very difficult to reach the extreme limits $V\to 0$ and $V\to \infty$ characterized by pure RKKY or IIME coupling, respectively, as the energy scale given by $T_{\rm C}$ becomes too small. 
As concerns the strong-coupling limit, we conclude that a perfect linear scaling of $T_{\rm C}$ with $\alpha$ can only be expected for still stronger hybridization strengths $V$ that are not accessible to DMFT with the presently used impurity solver.
This also implies that a direct comparison of the DMFT results for $T_{\rm C}$ with those obtained by static mean-field theory applied to the effective low-energy model is not meaningful. 
From Fig.\ \ref{fig:tc} we can infer that the latter would predict a Curie temperature which is by two orders of magnitude higher than the DMFT result for $V=7$ in Fig.\ \ref{fig:tcofv}. 
This could indicate that the strong-coupling limit is not yet reached but could also be ascribed to strong local fluctuations reducing $T_{\rm C}$ which are accounted for within the dynamical but not in the static mean-field theory. \cite{STK+06}

On the other hand, the maximum $T_{\rm C, opt} \approx 0.07$ found for intermediate $V$ is well accessible to DMFT and surprisingly high, in view of the fact that the magnetic coupling is mediated indirectly only. 
Compared to DMFT estimates \cite{KJMP05,Ulm98,MN00} of critical (N\'eel or Curie) temperatures in the Hubbard or periodic Anderson model with a dense system of correlated impurities, it is about an order of magnitude lower.
The optimal intermediate hybridization strength where $T_{\rm C}$ is at its maximum is given by $V_{\rm opt} \approx 2.5$, i.e., clearly stronger than the nearest-neighbor hopping $t=1$. 

One should note that DMFT applied the depleted Anderson lattice can be expected to be much more reliable than for the dense case. 
In fact, perfect agreement with numerically exact DMRG data is observed in the low-temperature limit. 
We are therefore convinced that this study provides quantitative insight into the physics and contributes to the fundamental understanding of magnetic order of correlated orbitals coupled indirectly by conduction electrons.

\acknowledgments

Support of this work by the Deutsche Forschungsgemeinschaft within the SFB 668 (project A14), by the excellence cluster ``The Hamburg Centre for Ultrafast Imaging -- Structure, Dynamics and Control of Matter at the Atomic Scale'' and by the SFB 925 (project B5) is gratefully acknowledged.

\appendix

\section{Low-temperature behavior of the magnetization at half-filing} 
\label{sec:app}

Here we derive the low-temperature behavior of $dm_{\rm A} / dT$ at half-filling, i.e., Eqs.\ (\ref{eq:m1}) and (\ref{eq:m2}).
Due to particle-hole symmetry at half-filling ($\mu = 0$), we have
\begin{equation}
  Q_\uparrow + Q_\downarrow = 2D
\end{equation}
and
\begin{equation}
  \eta_\uparrow({\bf k}) = - \eta_\downarrow({\bf k}) = -\frac{\alpha}{2} \left( 
  D-Q_{\downarrow}
  \right)
  \gamma^2({\bf k})
\end{equation}
Using Eq.\ (\ref{eq:occ}), we immediately have
\begin{equation}
  m_A
  =
  \frac{1}{L_A}\sum_{\ff k \in {\rm BZ}_{A}} 
  \tanh 
  \frac{
  \alpha(D-Q_\downarrow(T))
  \gamma^2({\bf k})
  }{
  4T
  } \, .
\end{equation}
In the thermodynamical limit, $L \mapsto \infty$, the $\bf k$-sum can be replaced by one-dimensional integration,
\begin{equation}
  m_A = \int_{-2D}^{2D} d\omega \rho^{(D)}(\omega) \tanh 
  \frac{\alpha(D-Q_\downarrow(T))\omega^2}{4T} 
  \, ,
\label{eq:mint}
\end{equation}
with the weight function
\begin{equation}
  \rho^{(D)}(\omega) = \frac{1}{L_A} \sum_{\bf k} \delta(\omega-\gamma({\bf k})) \, .
\end{equation}
We note that $\rho^{(1)}(\omega)$ and $\rho^{(3)}(\omega)$ are finite at the Fermi edge (at $\omega=0$), while $\rho^{(2)}(\omega) \sim \ln |\omega|$ diverges.
Since $\rho^{(D)}(-\omega) = \rho^{(D)}(\omega)$ and since $Q_\downarrow(T) \simeq 0$ for low temperatures, Eq.\ (\ref{eq:mint}) implies
\begin{eqnarray}
  \frac{d m_A}{dT} = -\frac{\alpha D}{2T^2} \int_0^{2D} d\omega \rho^{(D)}(\omega) \frac{\omega^2}{\cosh^2 \frac{\alpha D \omega^2}{4T}} \: .
\end{eqnarray}
After changing the integration variable, we obtain
\begin{equation}
  \frac{d m_A}{dT} \simeq -\frac{1}{\sqrt{\alpha D T}}\int_0^{\infty} dx \, \rho^{(D)}\left(2\sqrt{\frac{Tx}{\alpha D }} \right) \frac{\sqrt{x}}{\cosh^2 x}
\end{equation}
for low $T$.
Since $\rho^{(D)}(\omega)$ is regular at $\omega=0$ for $D=1$ and $D=3$, we have
\begin{eqnarray}
  \frac{d m_A}{dT}\Big|_{T=0} 
  \stackrel{T\to0}{\sim} 
  -\frac{1}{\sqrt{\alpha D T}}  \quad\quad\quad (D=1, D=3)
\end{eqnarray}
and thus Eq.\ (\ref{eq:m1}).
Furthermore, with $\rho^{(2)} ( \sqrt{2T/\alpha} ) \sim \ln (2T/\alpha)$ we get
\begin{eqnarray}
  \frac{d m_A}{dT} 
  \stackrel{T\to0}{\sim} 
  \frac{\ln (T/\alpha)}{\sqrt{\alpha D T}}  \quad\quad (D=2) \: ,
\end{eqnarray}
i.e., Eq.\ (\ref{eq:m2}).
The low-temperature behavior of $m_{\rm A}$ is thus governed by the weight function $\rho^{(D)}(\omega)$ at low $\omega$.
Analogously, one may also relate the low-temperature behavior of $m_{\rm A}$ to the van Hove singularity of the spin-dependent tight-binding density of states corresponding to the mean-field dispersion Eq.\ (\ref{eq:mfdis}).


\begin{thebibliography}{47}
\bibitem{ferro}
Strictly speaking, the ground state displays ferri- rather than ferromagnetic order since the local magnetic moments of the B sites are antiferromagnetically aligned to the magnetic moments of the A and of the impurity sites, see Fig.\ 2 of Ref.\ \onlinecite{STP13}, for example. This results from the effectively antiferromagnetic (Kondo-type) coupling of the impurity sites to the B sites. Since the absolute value of the B moment is small, we loosely denote the state as ``ferromagnetic'' in the rest of the paper.
\end{thebibliography}
\end{document}